\documentclass[twocolumn,showpacs,preprintnumbers,amsmath,amssymb, prb, reprint]{revtex4-1}

\usepackage{stmaryrd}
\usepackage{txfonts}
\usepackage{amssymb}
\usepackage{mathrsfs}
\usepackage{graphicx}
\usepackage{dcolumn}
\usepackage{bm}
\usepackage{epsfig}
\usepackage{color}                    
\usepackage[colorlinks=true,linkcolor=blue,citecolor=blue,urlcolor=blue]{hyperref}                 

\begin{document}
\preprint{\href{http://dx.doi.org/10.1103/PhysRevB.90.054509}{L. N. Bulaevskii and S.-Z. Lin, Phys. Rev. B {\bf{90}}, 054509 (2014).}}
\title{Dissipation in Josephson tunneling junctions at low temperatures }

\author{Lev N. Bulaevskii and Shi-Zeng Lin} 
\affiliation{Theoretical Division, Los Alamos National Laboratory, Los Alamos, New Mexico 87545, USA}

\date{\today}
 
\begin{abstract}
It is important to know the decoherence mechanism of a qubit based on Josephson junctions. At low temperatures, as quasiparticle concentration becomes exponentially small, one needs to consider energy transfer from  tunneling electrons to other degrees of freedom to find dissipation in Josephson junctions and decoherence in qubits. Here we discuss the energy transfer to two-level systems, i.e. the transitions between two different configurations of ions inside insulating layer separated by a potential barrier. We derive a general equation of motion for the phase difference between two superconducting electrodes and we find a retarded dissipation term  due to electromagnetic mechanism and also contribution due to electron tunneling mechanism. Using the equation of motion we calculate the decay of Rabi oscillations and frequency shift in qubits due to the presence of the two-level systems. In the long time limit our results coincide with those obtained by Martinis {\it et al.} [J. M. Martinis {\it et al.}. Phys. Rev. Lett. {\bf{95}}, 210503 (2005)] within the Fermi's Golden rule approach up to a numerical factor.
\end{abstract}
\pacs{74.78.-w, 74.50.+r, 85.25.Pb} 

\maketitle

\section{Introduction}
A Josephson tunneling junction can operate effectively as a phase qubit if one can achieve a long coherence time by reducing dissipation and the corresponding noise. \cite{JM, Makhlin,Martinis,Wal}  It is well established that dissipation in tunneling Josephson junctions  near the critical temperature $T_c$ and at intermediate temperature is caused by quasiparticles. \cite{Likharev,Kulik} However, at very low temperatures,
$T\ll 2\Delta_g$, the quasiparticles are frozen out and their contribution to dissipation is exponentially small
being proportional to $\exp(-2\Delta_g/T)$, where $\Delta_g$ is the superconducting gap. \cite{Eckern,Larkin} Here we have assumed that the electrodes are made of $s$-wave superconductors.  Hence,  another mechanism of irreversible energy transfer from tunneling Cooper pairs to other degrees of freedom should be at work as experimental data confirm the existence of dissipation and phase decoherence in qubits at very low temperatures.  \cite{Martinis,Chioresku,JM} We consider in the following only sources of dissipation intrinsic to Josephson junctions excluding external sources such as photon induced tunneling discussed in Ref.~\onlinecite{Pekola}. 
Dissipation due to excitation of phonons discussed in Refs.~\onlinecite{Helm,Ioffe,Ivanchenko,Maksimov} is not effective at very low energies and we will not account  for it here.

Following the discussions by Martinis {\it et al.}\cite{JM} we will consider the dissipation at low temperatures $T\ll\Delta_g$ originating from two-level systems, i.e. the transitions between two different configurations of ions inside amorphous  insulating layer separated by a potential barrier. It was well established experimentally that two-level systems are responsible for specific heat and ac dielectric losses at low temperatures and  frequencies below 20 GHz in amorphous dielectrics \cite{Hunklinger} which are inevitably present inside Josephson junctions. 

In Ref. \onlinecite{JM}, the decay of Rabi oscillations was obtained using the Fermi's Golden rule approach. The dynamics of Josephson junctions is governed by an equation of motion for the gauge invariant phase difference. Such an equation accounting for the presence of the two-level systems is not available to date. In the present work, we derive such an equation of motion by accounting for both the electromagnetic and tunneling mechanism for the dissipation caused by the two-level systems. The resulting general equation of motion can be used to describe the dynamics of junction at arbitrary time after the junction is perturbed away from equilibrium.

We will derive a general form of the dissipation term in the equation of motion for the phase difference in tunneling junctions with amorphous dielectric layer. We model this dielectric layer as an ensemble of two-level systems. Interaction of the phase difference with the two-level systems is electromagnetic in nature and results in retarded dissipation, i.e. the dissipative term in the equation for the phase difference is nonlocal in time and the corresponding equation for the phase difference is an integral one with respect to time. We also consider the tunneling mechanism for dissipation and show that it is less effective than the electromagnetic one. We then treat a qubit as a two-level system and use the Bloch equations for a ``spin" to describe the qubit and two-level systems in the insulating layer. We show that in the long time limit the dissipation of the low-amplitude Rabi oscillation in qubit differs from the results of Martinis {\it et al.}\cite{JM} only by a numerical factor and a weakly frequency-dependent logarithmic factor.

\section{Electromagnetic mechanism}

We consider low energy excitations inherent to two-level systems in amorphous dielectrics (i.e. SiO$_2$ and SiN$_x$) between junction electrodes. Specifically, oxygen ions in SiO$_2$ or nitrogen ions in SiN$_x$ may occupy two close positions in Si matrix separated by a distance $b_\alpha$ of the order of atomic length. In energy space these two configurations are separated by a potential barrier. We denote these positions as L and R, we take the energy of such states as $\epsilon_{LL}=\Delta_\alpha$ and $\epsilon_{RR}=-\Delta_\alpha$. In the space of states $|L\rangle$ and $|R\rangle$ tunneling of ion between these two configurations results in the off-diagonal matrix elements $\epsilon_{LR}=-\epsilon_{RL}=\Delta_{0,\alpha}$ in the Hamiltonian describing the two-level systems. The value of $\Delta_{0,\alpha}$ is related to the distance $b_\alpha$ between the two ions in the states $|L\rangle$ and $|R\rangle$ as
\begin{equation}
2\Delta_{0,\alpha}(b_\alpha)=\epsilon_a\exp(-b_\alpha/a), 
\end{equation}
where $\epsilon_a$ is of the order of atomic energy and $a$ is a characteristic atomic length. For a given junction, we assume that $\epsilon_a$ and $a$ are fixed while $b_\alpha$ varies from site to site with an index $\alpha$. In the presence of an electric field $E=\hbar\dot{\phi}/2ed$ between junction electrodes, the diagonal energies admit additional contributions $\pm e^*b_\alpha E\cos\eta_\alpha$ for $\epsilon_{LL}$ and $\epsilon_{RR}$, respectively. Here $\phi$ is the superconducting phase difference and $\dot{\phi}$ is its first time derivative. $d$ is the junction thickness, $e^*$ is the effective charge of tunneling ions and $\eta_\alpha$ is the angle between the vector ${\bf b}_\alpha$ and the direction of the electric field between electrodes, ${\bf z}$. These contributions due to the electric field describe the interaction ${\cal H}_{\mathrm{int}}$ between the two-level systems and the phase difference $\phi(t)$. The eigenstates of the two-level system are 
\begin{eqnarray}
&&|-\rangle=\sin(\theta_\alpha/2)|L\rangle-\cos(\theta_\alpha/2)|R\rangle, \\
&&|+\rangle=\sin(\theta_\alpha/2)|L\rangle+\cos(\theta_\alpha/2)|R\rangle, 
\end{eqnarray}
with the eigenvalues $\pm\hbar\Omega_\alpha/2\equiv \pm(\Delta_\alpha^2+\Delta_{0,\alpha}^2)^{1/2}$. Here $\tan\theta_{\alpha}=\Delta_{0,\alpha}/\Delta_\alpha$. 
The dipole moment of the two-level systems is $(e^*{\bf b}_{\alpha})\cos\theta_{\alpha}(|+\rangle-|-\rangle)$. The matrix element ${\cal H}_{\mathrm{int}}$ between states $|\pm\rangle$ is $\langle -|{\cal H}_{\mathrm{int}}|+\rangle=Ee^*b_\alpha\sin\theta_\alpha\cos\eta_\alpha$. The distribution of two-level system parameters $\Delta_\alpha$ and $b_\alpha$ is assumed to be uniform, 
\begin{equation}
P(\Delta_\alpha,b_\alpha)d\Delta_\alpha db_\alpha={\cal P}d\Delta_\alpha db_\alpha,\label{distr}
\end{equation}
where the normalization parameter ${\cal P}$ will be defined later. Such a tunneling model describes the experimental data on the specific heat and electric losses in amorphous dielectrics quite well up to frequencies about 20 GHz. \cite{Hunklinger} This frequency is much higher than the Josephson frequency and we take this frequency as a cutoff frequency in our theory. 
 In the $|\pm\rangle$ representation the Hamiltonian of interaction between the two-level system with the index $\alpha$ and the phase difference $\phi$ is 
\begin{eqnarray}
&&{\cal H}_s=\hbar[\Omega_{\alpha} \hat{S}_z+\lambda_{\alpha}\dot{\phi}(\cot\theta_{\alpha}\hat{S}_z+\hat{S}_x)], \label{EM}\\&&\lambda_{\alpha}=(2e^*b_{\alpha}/ed)\sin\theta_{\alpha}\cos\eta_{\alpha},
\end{eqnarray}
where $\hat{S}_k=\hat{\sigma}_k/2$ and $\hat{\sigma}_k$ are Pauli matrices with $k=x,\ y,\ z$. The junction Hamiltonian is
\begin{equation}
{\cal H}_J=J_0A[1-\cos\phi+\omega_J^{-2}\dot{\phi}^2/2- I \phi],
\end{equation}
where $J_0$ is the Josephson coupling density, $A$ is the junction area, $\omega_J$ is the Josephson frequency and $I$ is the bias current $I_B$ via the junction in units of the Josephson critical current $I_c$, $I\equiv I_B/I_c$.

Using the Heisenberg equation, we obtain the Bloch equations for ``spin" variables ${\bf S}_{\alpha}$
\begin{eqnarray}
&&\dot{S}_{x,\alpha}=-(\Omega_{\alpha}+\lambda_{\alpha}\dot{\phi}\cot\theta_{\alpha}) S_{y,\alpha}, \\
&&\dot{S}_{y,\alpha}=(\Omega_{\alpha} +\lambda_{\alpha}\dot{\phi}\cot\theta_{\alpha})S_{x,\alpha}-\lambda_{\alpha}\dot{\phi}S_{z,\alpha}.
\end{eqnarray}
The solution in the case of weak coupling, i.e. small $S_{x,\alpha},\ S_{y,\alpha}\ll 1$ and $S_{z,\alpha}\approx 1$ when $\lambda_\alpha\dot{\phi}\ll \Omega_\alpha$, is 
\begin{eqnarray}
&&S_{x,\alpha}=\lambda_{\alpha}\int_0^tdt'\dot{\phi}(t')\sin[\Omega_{\alpha}(t-t')], \\
&&\dot{S}_{x,\alpha}=\lambda_{\alpha}\Omega_{\alpha}\int_0^tdt'\dot{\phi}(t')\cos[\Omega_{\alpha}(t-t')],
\nonumber
\end{eqnarray}
and the two-level system frequency is renormalized from $\Omega_{\alpha}$ to $\Omega_{\alpha}+\dot{\phi}\lambda_{\alpha}\cot\theta_{\alpha}$. We replace $\Omega_{\alpha}+\dot{\phi}\lambda_{\alpha}\cot\theta_{\alpha}$ by $\Omega_\alpha$ in the following discussions. Initially the junction is assumed in equilibrium, $\dot{\phi}(t=0)=0$.
In the equation for the phase difference we obtain the dissipation contribution due to the two-level systems:
\begin{equation}\label{clas}
\omega_J^{-2}\ddot{\phi}+\sin\phi-\sum_{\alpha}(\hbar\lambda_{\alpha}/J_0A)\dot{S}_{x,\alpha}=I.
\end{equation}
We replace summation over $\alpha$ by integration over $\Delta$ and $b$ with a uniform distribution function Eq. \eqref{distr}. In the integral over $\Delta$ and $b$ we replace variables $\Delta$ and $b$ by new variables $\hbar\Omega=(\Delta^2+\Delta_0^2)^{1/2}$ and $\sin\theta=\Delta_0/(\hbar\Omega)$:
\begin{eqnarray}
&&d\Delta db=-\frac{a }{\Delta_0}d\Delta d\Delta_0=\frac{a\hbar }{\Delta_0||W||}d\Omega d(\sin\theta), \\
&&||W||=\hbar(\partial_{\Delta}\Omega\partial_{\Delta_0}\sin\theta-\partial_{\Delta_0}\Omega\partial_{\Delta}\sin\theta)=\cos\theta /(\hbar\Omega), 
\end{eqnarray}
The last term in the left hand side of Eq.~\eqref{clas} gives the dissipation term
\begin{eqnarray}
&&\frac{2\hbar \mu\sigma d}{J_0\Omega_m}\int_0^{\Omega_m}d\Omega\Omega d(\sin\theta)d\eta\tan\theta\ln^2\left(\frac{\epsilon_a}{\hbar\Omega\sin\theta}\right)\cos^2\eta\times \nonumber\\
&&\int_0^tdt'\dot{\phi}(t')\cos[\Omega(t-t')] \nonumber\\
&&\approx \int_0^{\Omega_m}\frac{d\Omega}{\Omega_m}{\cal E}(\Omega)\int_0^tdt'\dot{\phi}(t')\cos[\Omega(t-t')], \label{tteq14}\\
&&{\cal E}(\Omega)=\frac{\pi \hbar \mu\sigma \Omega d}{2J_0}\ln^2\left(\frac{\Omega_a}{\Omega}\right)
=\frac{\Omega\Gamma_0}{\omega_J^2}\ln^2\left(\frac{\Omega_a}{\Omega}\right),\nonumber\\
&&\Gamma_0=
\frac{2\pi^2\mu\sigma d^2e^2}{\epsilon_d\hbar}, \ \ \ \Omega_a=\frac{\epsilon_a}{\hbar},
\end{eqnarray}
where $\mu=2e^*a/ed$,  while the cutoff frequency is $\Omega_m\approx 20$ GHz and $\sigma$ is the volume density of the two-level systems.  Here $\epsilon_d$ is the high frequency dielectric function. In obtaining Eq. \eqref{tteq14}, we have neglected $\sin\theta$ inside logarithm. It only results in a numerical factor of the order of unity inside the logarithm, $\ln^2(\epsilon_a/\hbar\Omega)$, which is omitted within logarithmic accuracy. The logarithmic factor $\ln(\Omega_a/\Omega)$ is valid only for $\Omega>\omega_0\approx \Omega_a\exp(-1/\sigma^{1/3}a)$ as we have assumed that the two-level systems do not overlap. At lower frequencies, $\Omega$ inside the logarithmic factor should be replaced by $\omega_0$.

\section{Tunneling mechanism}

Displacements of ions described by the two-level system model induce the electric field inside the dielectric and thus change the electric potential $V$ barrier for tunneling electron. The electron tunneling integral $\beta$ is given as
\begin{equation}
\beta=\epsilon_a\exp\left[-\hbar^{-1}\int_0^d dz(2m_e[eV(z)-\epsilon_e])^{1/2}\right],
\end{equation}
where $m_e$ is the electron mass and $\epsilon_e$ is its energy. We consider the case that the frequencies $\Omega$ of two-level systems are much lower than the inverse transversal time for tunneling, $1/\tau_T$, where $\tau_T$ is given by the quasiclassical formula 
\begin{equation}
\tau_T=\int_0^d dz\sqrt{\frac{m_e}{2[eV(z)-\epsilon_e]}}.
\end{equation}
The change of $\beta $ due to ion displacements, $\delta \beta=\beta-\beta_0$, is
\begin{equation}
\frac{\delta \beta}{\beta_0}=\frac{m_eed}{\hbar^2\ln(\epsilon_a/\beta_0)}\int_0^d dz\delta V(z)=\frac{\ln(\epsilon_a/\beta_0)}{V_0d}\int_0^d dz\delta V(z),
\end{equation}
where $V_0\approx e/a$. The change of  potential, $\delta V(z)$, induced by the dipoles ${\bf p}_{\alpha}=e^*{\bf b}_{\alpha}$ of two-level systems positioned at ${\bf r}_{\alpha},\ z_{\alpha}$, with ${\bf r}_{\alpha}=(x_{\alpha},\ y_{\alpha})$, is
\begin{equation}
\delta V(z)=\frac{e^*}{\epsilon_d}({\bf b}\cdot\nabla)\left[\frac{1}{[({\bf r}-{\bf r}_{\alpha})^2+(z-z_{\alpha})^2]^{1/2}}\right].
\end{equation} 
The change in the Josephson coupling density $\delta J_0$ caused by the change in electron tunneling $\delta\beta$ is $\delta J_0/J_0\approx 2\delta \beta/\beta_0$. For the tunneling mechanism the Hamiltonian for the interaction between the two-level systems with index $\alpha$ and the phase difference is
\begin{eqnarray}
&&{\cal H}_{T,\alpha}=-J_0\int d{\bf r}\frac{\ln(\epsilon_a/\beta_0)}{V_0d}(1-\cos\phi)\hat{S}_{x}\int_0^d dz\delta V({\bf r},z)= \nonumber\\
&&J_0A\frac{\ln(\epsilon_a/\beta_0)}{V_0d}(1-\cos\phi)\hat{S}_{x}\int_0^d dz\delta V({\bf r}=0,z).\label{T}
\end{eqnarray}
Using the Bloch equations for a ``spins" we find
\begin{eqnarray}
&&S_{x,\alpha}(t)=-\Omega_{\alpha}{\cal B}_{\alpha}\int_0^t dt'[1-\cos\phi(t')]\sin[\Omega_{\alpha}(t-t')]=\nonumber\\
&&-{\cal B}_{\alpha}\int_0^t dt'\sin\phi(t')\cos[\Omega_{\alpha}(t-t')], \\
&&{\cal B}_{\alpha}=\int_0^d dz[b_{z,\alpha}(z-z_{\alpha})-(b_{x,\alpha}x_{\alpha}-b_{y,\alpha}y_{\alpha})]R^{-3}(z), \nonumber\\
&&R^2(z)=(z-z_{\alpha})^2+{\bf r}_{\alpha}^2.
\end{eqnarray}
The coordinates ${\bf r}_{\alpha},\ z_{\alpha}$ are defined up to the size of the two-level system dipole $b$. In the equation for the phase difference, Eq. \eqref{clas}, we have an additional two-level system contribution due to the tunneling mechanism
\begin{eqnarray}
&&\left[\frac{e^*\ln(\epsilon_a/\beta_0)}{\epsilon_dV_0d}\right]^2\sum_{\alpha}{\cal A}_{\alpha}\sin\phi(t)\int_0^tdt'\sin\phi(t')\cos[\Omega(t-t')],\nonumber\\
&&{\cal A}_{\alpha}=\int_0^ddu\int_0^ddv\frac{b_{z,\alpha}^2(u-z_{\alpha})(v-z_{\alpha})+(b_{x,\alpha}x_{\alpha}+b_{y,\alpha}y_{\alpha})^2}{R^3(u)R^3(v)}.\nonumber
\end{eqnarray}
We replace the summation $\sum_\alpha$ by integration over coordinates ${\bf r}_{\alpha},\ z_{\alpha}$ of two-level systems
\begin{equation}
{\cal A}=\sum_{\alpha}{\cal A}_{\alpha}\Rightarrow\sigma\int d{\bf r}_{\alpha}\int_0^d dz_{\alpha}{\cal A}({\bf r}_{\alpha},z_{\alpha}).
\end{equation}
Integration over ${\bf r}_{\alpha}$ and averaging over directions of two-level system dipoles ${\bf b}_\alpha$ gives
\begin{eqnarray}
&&{\cal A}=\frac{\pi b^2\sigma}{2}\int_0^ddz\int_{-z}^{d-z}\int_{-z}^{d-z}dudv\left[\frac{2uv}{|u||v|(|u|+|v|)^2}+ \right.\nonumber\\
&&\left.\frac{\sqrt{2}(u^4+v^4)^{1/2}}{(u^2+v^2)^2+\sqrt{2}(u^2+v^2)(u^4+v^4)^{1/2}}\right]\approx \nonumber\\
&&\pi^2b^2\sigma d\ln\left(\frac{d}{b}\right).
\end{eqnarray}
We estimate the contribution  to the dissipation due to the tunneling mechanism in the equation for the phase difference, Eq. \eqref{clas}, as
\begin{eqnarray}
&&\int_0^{\Omega_m}\frac{d\Omega}{\Omega_m}{\cal T}(\Omega)\sin\phi(t)\int_0^tdt'\dot{\phi}(t')\sin\phi(t')\cos[\Omega(t-t')].\nonumber\\
&&{\cal T}(\Omega)=\left(\frac{\pi e^*\ln(\epsilon_a/\beta_0)}{\epsilon_de}\right)^2\frac{a^4\sigma\ln(d/a)}{d}\ln^2\left(\frac{\Omega_a}{\Omega}\right).
\end{eqnarray}
The ratio $r$ of the tunneling contribution to the dissipation and that of the electromagnetic mechanism at a given $\Omega$ and small $\phi$ is
\begin{equation}
r\approx\phi^2\frac{\pi^2\ln(\epsilon_\alpha/\beta_0)}{\epsilon_d^2}\left(\frac{a}{d}\right)\left(\frac{\hbar d \omega_J^2}{e^2\Omega}\right)\ln\left(\frac{d}{a}\right).
\end{equation}
Due to $a\ll d$ and $\omega_J\ll e^2/(\hbar d)$ the tunneling mechanism contribution is small everywhere except at low $\Omega$, where it remains nonzero in the limit $\Omega\rightarrow 0$, while the electromagnetic contribution vanishes in this limit. The different behavior for both mechanisms in this limit is because the two-level system interacts directly with the phase difference via $\cos\phi$
for the tunneling mechanism as shown in Eq.~\eqref{T}, while in the electromagnetic mechanism it interacts with $\dot{\phi}$ as shown in Eq.~\eqref{EM}.
We note that the tunneling mechanism gives contribution to the dissipation which is nonlinear in the phase difference and thus it is negligible at small deviations from the equilibrium.

\section{Decay of Rabi oscillations in the absence of bias current}

We consider dissipation for small-amplitude oscillations at $I_B=0$ neglecting the tunneling contribution.  We need to solve the equation
\begin{eqnarray}
&&\omega_J^{-2}\ddot{\phi}+\phi+ \nonumber\\
&&\int_0^{\Omega_m}\frac{d\Omega}{\Omega_m}{\cal E}(\Omega)\int_0^tdt'\dot{\phi}(t')\cos[\Omega(t-t')]=0.
\end{eqnarray}
Note that if ${\cal E}$ is frequency independent and $\Omega_m$ is infinite, we would obtain for the last term in the left-hand side the standard dissipation term $\tilde{\gamma}\dot{\phi}$. 

Neglecting the logarithmic factor in ${\cal E}(\Omega)$, we integrate over $\Omega$. After changing variables 
$t-t'=u/\Omega_m$, we get
\begin{eqnarray}
&&\omega_J^{-2}\ddot{\phi}+\phi+\frac{\Gamma_0}{\omega_J^2}\int_0^{\Omega_mt}du\dot{\phi}(t-u/\Omega_m)f(u)=0,
\label{c}\\
&&f(u)=\sin u/u-(1-\cos u)/u^2.\nonumber
\end{eqnarray}
To understand the long-time behavior we take $\phi(t)=\exp(i\omega t)$ in Eq.~(\ref{c}) and obtain for the left-hand side
\begin{eqnarray}
\left[-\omega_J^{-2}\omega^2\phi+\phi+\frac{i\omega\Gamma_0}{\omega_J^2}\int_0^{\Omega_mt}du
\exp(-i\omega u/\Omega_m)f(u)\right]\exp(i\omega t).\nonumber
\end{eqnarray}
We see that $\phi(t)=\exp(i\omega t)$ is indeed the solution of Eq.~(\ref{c}) in the long time limit $\Omega_mt\gg1 $ if we take the complex number $\omega$ as $\omega=\omega_J+\delta\omega+i\gamma$ with
\begin{equation}
\gamma\approx \frac{\pi\omega_J\Gamma_0}{4\Omega_m}, \ \ \ \ \ \delta\omega\approx \frac{\Gamma_0\omega_J}{2\Omega_m}\ln\left(\frac{\Omega_m}{\omega_J}\right)
\end{equation} 
for $\omega_J\ll\Omega_m$. The parameter $\gamma$ is the dissipation rate and $\delta\omega$ is the shift of Rabi oscillation frequency due to interaction with two-level systems. Accounting for the logarithmic factor, we obtain
\begin{eqnarray}
&&\gamma\approx\frac{\pi\omega_J\Gamma_0}{4\Omega_m}\ln^2\left(\frac{\Omega_a}{\Omega_m}\right),\\ 
&&\delta\omega\approx \frac{\Gamma_0\omega_J}{2\Omega_m}\ln\left(\frac{\Omega_m}{\omega_J}\right)\ln^2\left(\frac{\Omega_a}{\Omega_m}\right). \label{cl2}
\end{eqnarray} 
Thus the retardation effect is not important at $\Omega_m\gg\omega_J$ in the stationary solution which is established in the long-time limit $\Omega_mt\gg 1$.  \\

\section{Dissipation of Rabi oscillations in a qubit}

We consider now a Josephson junction in the presence of a bias current. Then energy levels in the quantum regime become non-equidistant and we consider the operation only between the ground state and the first excited state by ignoring the other energy levels valid for a sufficiently high bias current $I_B$. We replace $\dot{\phi}$ by the momentum operator and the Hamiltonian for a qubit takes the form
\begin{equation}
{\cal H}_J=J_0A(1-\cos\phi)-4E_c\partial^2/\partial\phi^2-\hbar I_B\phi/(2e),
\end{equation}
where  $E_c=e^2/2C$ and $C$ is the junction capacitance. The bias current consists of a dc and an ac component, $I_B=I_{ac}+I_{dc}$. We find the Hamiltonian of interaction between the oscillator and two-level systems by replacing $\dot{\phi}$ by the momentum operator $\hat{P}=i\partial/\partial\phi$ according to the relation $\dot{\phi}\rightarrow 8\hat{P}E_c/\hbar$. Then the qubit is described by a two-level system with the Pauli matrices $\hat{{\bf Q}}$ for a qubit ``spin". The matrix element of the operator $\hat{P}$ between the ground state and the first excited state of the oscillator is given by the expression $i(\hbar\omega_{10}e^2/2C)^{1/2}$ and we write the total Hamiltonian for the qubit and the two-level systems as
\begin{eqnarray}
&&{\cal H}=\hbar\left[\frac{\omega_{10}}{2}\hat{Q}_z-I_{ac}\hat{Q}_y/(2e)+\sum_{\alpha}\frac{\Omega_{\alpha}}{2}\hat{S}_{z,\alpha}\right]+{\cal H}_{\mathrm{int}}, \\
&&{\cal H}_{\mathrm{int}}=\frac{1}{2}\sum_{\alpha}P_{\alpha}\hat{Q}_x\hat{S}_{x,\alpha}, \ \ \ \ P_{\alpha}=i\frac{4b_{z,\alpha}}{ d}\left(\frac{\hbar\omega_{10}ee^*}{2C}\right)^{1/2},
\end{eqnarray}
where $\omega_{10}$ is the energy difference between the first excited state and the ground state of the qubit. This Hamiltonian describes the transfer of energy from the qubit to the two-level systems, and then back from the two-level systems to the qubit. Experimentally, a coherent state of qubit oscillation between the ground and the first excited state is prepared at time $t=0$ and then probability to find the qubit in the excited state at time $t$ is measured. Using the Heisenberg equation of motion for ``spin" operators, we obtain the Bloch equations for the system
\begin{eqnarray}
&&\dot{Q}_x=-\omega_{10}Q_y-I_{ac}Q_z/e, \ \ \ \ \ \dot{Q}_y=\omega_{10}Q_x-\sum_{\alpha}P_{\alpha}S_{x,\alpha}Q_z/\hbar, \nonumber\\
&&\dot{S}_{x,\alpha}=-\Omega_{\alpha}S_{y,\alpha}, \ \ \ \ \ \dot{S}_{y,\alpha}=\Omega_{\alpha}S_{x, \alpha}+P_{\alpha}S_{z,\alpha}Q_x/\hbar. \label{gen}
\end{eqnarray}
These equations describe the dynamics of a qubit in the presence of $I_{ac}$ and two-level systems in the insulating layer. For weak excitations, $Q_x,\ Q_y\ll1$, $S_{x,\alpha},\ S_{y,\alpha}\ll1$ and $Q_z\approx 1$, $S_{z,\alpha}\approx 1$ at $I_{ac}=0$, after excluding the operators for the two-level systems, we obtain the equation for the qubit 
\begin{eqnarray}
\ddot{Q}_x+\omega_{10}^2Q_x+\sum_{\alpha}\frac{|P_{\alpha}|^2}{\hbar^2}
\int_0^tdt' Q_x(t')\sin[\Omega_\alpha(t-t')]=0. \nonumber
\end{eqnarray}
After summation over all two-level systems, this equation within a logarithmic accuracy takes the form
\begin{eqnarray}
&&\omega_{10}^{-2}\ddot{Q}_x+Q_x+\nonumber\\
&&\Gamma_0\ln^2\left(\frac{\Omega_a}{\Omega_m}\right)\int_0^tdt'  
Q_x(t')\frac{1-\cos[\Omega_m(t-t')]}{\Omega_m(t-t')}=0. \label{tteq36}
\end{eqnarray}
We assume that the qubit is perturbed away from equilibrium state at time $t=0$, i.e. $Q_x(t=0)\neq 0$. For time $t<\Omega_m^{-1}$ dissipation is absent due to its retarded nature. In the stationary state at time $t\gg \Omega_m^{-1}$, the retardation becomes ineffective and we take the solution as $Q_x(t)\sim \exp(i\omega t)$ with $\omega=\omega_{10}+\delta\omega+i\gamma$ determined by
\begin{eqnarray}
&&1-\frac{\omega^2}{\omega_{10}^2}+ 
\frac{\Gamma_0}{\Omega_m}\ln^2\left(\frac{\Omega_a}{\Omega_m}\right)\int_0^{\infty}du \exp\left[\frac{i\omega u}{\Omega_m}\right]  g(u)=0, \\
&&g(u)=(1-\cos u)/u.\nonumber
\end{eqnarray}
We then obtain for the dissipation rate and the frequency shift
\begin{eqnarray}
&&\gamma\approx\frac{\pi\omega_{10}\Gamma_0}{2\Omega_m}\ln^2\left(\frac{\Omega_a}{\Omega_m}\right).\\
&&\delta\omega\approx\frac{\omega_{10}\Gamma_0}{\Omega_m}\ln\left(\frac{\Omega_m}{\omega_{10}}\right)\ln^2\left(\frac{\Omega_a}{\Omega_m}\right).
\end{eqnarray}
At intermediate time $t\sim \Omega_m^{-1}$, one needs to solve Eq. \eqref{tteq36}. 

Result for $\gamma$ coincides with 
that of Martinis {\it et al.} \cite{JM}, $\gamma\approx{\pi\omega_{10}\Gamma_0}/{(6\Omega_m)}$,  except for a numerical factor  and the  logarithmic factor $\ln^2(\Omega_a/\Omega_m)$ missed in their treatment. 
In our approach, as well as that of Ref.~\onlinecite{JM}, transfer of energy to noninteracting two-level systems was assumed, i.e. the dipole-dipole interaction of two-level systems was neglected. As shown in Ref.~\onlinecite{Burin}, this is possible if the amplitude of Rabi oscillations, $Q_x(t=0)$, is small, i.e. $(4\pi\sigma/\hbar \Omega_m)(e^*a)^2\sqrt{Q_x(t=0)}\ll 1$. For $\sigma/\hbar\Omega_m\approx 3\times 10^{29}$\ $\mathrm{(erg\cdot cm^3)^{-1}}$, see Ref.~\onlinecite{JM}, this condition is fulfilled very well.

Excitation of Rabi oscillations by an ac current is described by Eqs.~(\ref{gen}). For short external current pulses 
with duration $\tau$, the retarded dissipation due to two-level systems may be neglected at any current if $\Omega_m\tau\ll 1$.
Otherwise, the condition of ignoring dissipation is $I_{dc}/e\gg \gamma$. For stationary solutions, the retarded nature of dissipation term is not important.

\section{Conclusions}

In conclusion, we have derived a general form of two-level systems driven dissipative terms in equation of motion for the phase difference in tunneling junctions with an amorphous dielectric layer at very low temperatures when dissipation caused by quasiparticles is negligible. We account for the direct electromagnetic mechanisms of phase and two-level systems interaction and also for the effect of two-level systems on electron tunneling in junctions. We show that they give terms which are nonlinear and linear in phase, respectively.  We find that the dissipation from the tunneling mechanism is less effective than the electromagnetic one. Finally we have derived the decay rate of the Rabi oscillations due to the presence of two-level systems in the insulating layer of Josephson junctions. Our results are consistent with those obtained in Ref. \onlinecite{JM}  up to a numerical factor and a weakly frequency-dependent logarithmic factor.
\section{ACKNOWLEDGMENTS}

We acknowledge helpful discussions with V. B. Geshkenbein, G. Blatter, J. Martinis, L. Ioffe, A. Golubov, A. Ustinov,  V. Ryazanov 
and D. Khmel'nitskii. L.N.B. thanks the Pauli Cenetr, ETH, Zurich for hospitality. The work was carried out under the auspices of the NNSA of the US DOE at LANL under Contract No. DE-AC52-06NA25396.

\end{document}